\documentclass[twocolumn]{pasj00}
\begin{document}
\SetRunningHead{T. Sonoi and H. Shibahashi}{TDC analysis of strange-modes in hot massive stars}
%\Received{}%{yyyy/mm/dd}
%\Accepted{}%{yyyy/mm/dd}
%\Published{}%{yyyy/mm/dd}

\title{Stability Analysis of Strange-Modes in Hot Massive Stars \\ with Time-Dependent Convection}

%%% begin:list of authors
% Do NOT capitalize all letters in "textsc".
\author{Takafumi \textsc{Sonoi}}
\affil{LESIA, Observatoire de Paris, 5 Place Jules Janssen, 92195, Meudon Cedex, France}
\email{takafumi.sonoi@obspm.fr}
\author{and}
\author{Hiromoto \textsc{Shibahashi}}
\affil{Department of Astronomy, School of Science, The University of Tokyo, Hongo 7-3-1, Bunkyo-ku, Tokyo 113-0033}
\email{shibahashi@astron.s.u-tokyo.ac.jp}
%%% end:list of authors

%%% Please use the following style in case that sorting by 
%%% affiliation is impossible. 
%
% \author{%
%   D-Firstname \textsc{D-Familyname}\altaffilmark{1}
%   E-Firstname \textsc{E-Familyname}\altaffilmark{1,2}
%   and
%   F-Firstname \textsc{F-Familyname}\altaffilmark{2}}
% \altaffiltext{1}{Address of Institute}
% \email{ddddd@xxx.xxx.xx.xx}
% \email{eeeee@xxx.xxx.xx.xx}
% \altaffiltext{2}{Address of Institute}

%% `\KeyWords{}' always has to be placed before `\maketitle'.
 %Do NOT move this preamble from here!

\maketitle

\begin{abstract}
We carry out a nonadiabatic analysis of strange-modes in hot massive stars with time-dependent convection (TDC). In envelopes of such stars, convective luminosity is not so dominant as that in envelopes of stars in the redder side of the classical instability strip. Around the Fe opacity bump, however, convection contributes non-negligibly to energy transfer. Indeed, we find that instability of modes excited at the Fe bump is likely to be weaker with TDC compared with the case of adopting the frozen-in convection approximation. But we confirm that unstable strange-modes certainly remain in hot massive stars even by taking into account TDC. We also examine properties of the strange-mode instability, which is related to destabilization of strange-modes without adiabatic counterparts. In this type of instability, the phase lag between density and pressure varies from 0 to $180^{\circ}$ in an excitation zone unlike the case of the $\kappa$-mechanism. In addition, we confirm by comparing models with $Z=0$ and $Z=0.02$ that dominance of radiation pressure is important for this type of instability.
\end{abstract}

\section{Introduction}
Strange-modes are one type of stellar pulsation modes, but have significantly different properties from those of ordinary modes appearing in most of pulsating stars. Although their physical properties have not been well established yet, they have been examined by many authors so far. \citet{Wood1976} found strange-modes for the first time in a numerical study for luminous helium stars. He pointed out that there is no one-to-one correspondence between solutions by adiabatic and nonadiabatic analyses. As a matter of fact, they were not called strange-modes at that time, but \citet{Cox1980} named them ``strange'' modes for the first time in the study of pulsations in hydrogen deficit carbon stars. \citet{Shibahashi1981} systematically analyzed radial and nonradial pulsations in models with different $L/M$ ratios, and found that strange-modes appear in models with $L/M\gtrsim 10^4L_{\odot}/M_{\odot}$. They also investigated the origin of strange-modes with a numerical experiment, in which they artificially changed the thermal time-scale. For the thermal time-scale reduced to zero, which generates an extremely nonadiabatic situation, the strange-mode eigenfrequencies are close to those obtained by a fully nonadiabatic analysis with the realistic and unchanged thermal time-scale. On the other hand, when increasing the thermal time-scale and generating an adiabatic situation, the real part of eigenfrequency decreases toward zero while the imaginary part remains large. And then the mode becomes an oscillatory convection ($g^{-}$) mode. \citet{Saio1984} performed a similar experiment, and found a relation to thermal waves.

\citet{Gautschy1990} found strange-modes not related to thermal waves both in a fully nonadiabatic analysis and in the nonadiabatic reversible (NAR) approximation (equivalent with reducing the thermal time-scale to zero in \authorcite{Shibahashi1981}'s experiment). In the NAR approximation, the classical $\kappa$-mechanism can no longer work, and hence an alternative physical explanation for the excitation of the strange-modes has been needed. To understand about the instability of this type of strange-mode, which is called ``strange-mode instability'', \citet{Glatzel1994} suggested with a local analysis that dominance of radiation pressure leads to a large phase lag between density and pressure perturbations, and to the strange-mode instability. \citet{Saio1998} also carried out analytic investigations through different approaches, and obtained similar consequences. But in addition to this, they claimed that the opacity derivative with respect to density $\kappa_\rho$ is essential for the instability, and that the instability grows as radiation pressure gradient produces a restoring force. 

Since the strange-modes mentioned above appear in the environment with short thermal time-scale, the adiabatic approximation is no longer available for them. However, after new opacity tables \citep{Rogers1992} were released, strange modal sequences have been found even by adiabatic analyses in a modal diagram, i.e. a diagram plotting frequencies as a function of a stellar parameter (e.g. mass, effective temperature). Modes on such sequences appear since the Fe opacity bump, enhanced in the new opacity tables, causes a sound speed inversion. As a result, mode amplitude is confined around there, and the $\kappa$-mechanism works efficiently and leads to an extremely rapid growth of amplitude (Kiriakidis et al. \yearcite{Kiriakidis1993}; Saio et al. \yearcite{Saio1998}).  

Although the physical properties still remain puzzling, unstable strange-modes have been found in models of very luminous stars such as massive stars, Wolf-Rayet stars, helium stars, etc. by many studies (summarized by Saio et al. \yearcite{Saio1998}). Growth time-scales of unstable strange-modes are likely to be much shorter than those of ordinary modes, and comparable to their pulsational periods. Then, instability of the strange-modes might lead to nonlinear phenomena such as mass loss, and might be influential on stellar evolution. Indeed, the instability of strange-modes has been suggested as one of the candidates for a trigger of the \authorcite{Humphreys1979} (HD, 1979) limit phenomenon. In the Hertzsprung-Russel (HR) diagram, there are few observed stars over the HD limit. This implies that stars with $M\gtrsim 50M_{\odot}$ cannot evolve toward red supergiants. Just in the lower left side of the HD limit, luminous blue variables [LBVs or S Dor (SD) variables] are distributed. They intermittently show irregular variations in visible magnitude in a timescale of years to decades, while their bolometric magnitude almost keeps constant. In the HR diagram, this phenomenon can be shown as a horizontal transition. As an explanation for this, it is thought that sporadic eruptions would take place by some mechanism, and generate a thick envelope with a pseudo-photosphere. As the envelope expands, the apparent effective temperature decreases. When the eruptions cease, the core would be exposed again, and the effective temperature goes back to the original high value. After repeating this process and losing substantial mass, the stars are thought to evolve toward Wolf-Rayet stars. 

Although the mass-loss mechanism of LBVs has not been established yet, Kiriakidis, Fricke \& Glatzel (\yearcite{Kiriakidis1993}) found that strange-modes are unstable around the HD limit, and suggested that their instability could be responsible for the HD limit phenomenon. Recently, nonlinear calculations for radial strange-modes have been carried out (\cite{Dorfi2000}; Chernigovski et al. \yearcite{Chernigovski2004}; Grott et al. \yearcite{Grott2005}) to investigate pulsationally-driven mass-loss, although there seem to be no conclusive results so far. On the other hand, \citet{Aerts2010} observed a pulsation in a luminous B star, HD 50064, and found that its mass-loss rate changes in a time-scale of the pulsation period. \citet{Godart2011} suggested that a strange-mode could be a candidate for this pulsation in terms of the period. 

Stability of strange-modes has been so far investigated by nonadiabatic analyses with frozen-in convection (FC) approximation. Indeed, convection theories still have a lot of uncertainties. Moreover, convective energy transport in the envelopes of hot massive stars is not so dominant as that of stars in the redder side of the classical instability strip. However, around the Fe opacity bump, convective luminosity occupies a few dozen percent of luminosity. Therefore, we cannot definitively conclude that convection never affects pulsations. Note that \citet{Glatzel1996} performed nonadiabatic analyses by two types of FC with zero Lagrangian and Eulerian perturbations of convective luminosity, and obtained significantly different results between the two types of FC. 

In this study, we carry out a nonadiabatic analysis of the strange-modes in hot massive stars with time-dependent convection (TDC). Despite uncertainties of convection theories, nonadiabatic analyses with TDC have been able to roughly explain suppression of pulsational instability in the redder side of the classical instability strip \citep{Baker1979, Gonczi1980, Gonczi1981, Houdek2000, Xiong2001, Dupret2005}. Convection in the redder side of the strip is caused by the H and the He opacity bumps. In hot massive stars, on the other hand, the Fe bump generates a convection zone with a certain contribution of convective luminosity. Pulsations in massive stars are then worthy to analyze with TDC. 
 
\section{Procedures}
We construct evolutionary models of massive stars with Modules for Experiments in Stellar Astrophysics (MESA, \authorcite{Paxton2011} \yearcite{Paxton2011}, \yearcite{Paxton2013}). The mixing length parameter is set to $\alpha=2.0$. No mass loss is taken into account. The atmosphere part is constructed by following the Eddington-grey recipe suggested by \citet{Paczynski1969}. 
We adopt the heavy element abundance ratios of \citet{Asplund2009} to evaluate opacity.

We carry out a stability analysis of radial pulsations with the nonadiabatic code developed by \citet{Sonoi2012}. In this study, perturbation of nuclear energy generation rate $\varepsilon$ is neglected. The outer boundary conditions are imposed at $\tau=10^{-4}$. We adopt the TDC formulation originally derived by \citet{Unno1967}. This theory has been further developed by \authorcite{Gabriel1974} (\yearcite{Gabriel1974}, \yearcite{Gabriel1975}) and \authorcite{Gabriel1987} (\yearcite{Gabriel1987}, \yearcite{Gabriel1996}, \yearcite{Gabriel1998}, \yearcite{Gabriel2000}). Recently, \citet{Grigahcene2005} independently implemented this theory into a nonadiabatic code, with which \citet{Dupret2005} succeeded in explaining the suppression of the pulsational instability of $\delta$ Scuti in the redder side of the classical instability strip. In this study, the perturbation of the convective flux is taken into account, while the effect of the Reynolds stress tensor perturbation is neglected. We set the parameter for this TDC theory to $\beta=1$ following \citet{Grigahcene2005} and \citet{Dupret2005}.
%%% ADDED BY FOLLOWING THE REFEREE %%%
It is still uncertain about how to deal with this parameter. The result of the stability is slightly dependent on the value of $\beta$. If we adopt different values for the parameter $\beta$, stability (or instability) of modes with small growth or damping rates changes to some extent, but the results for those with large rates do not change significantly.
%%%%%%%%%%%%%%%%%%%%%%%%%%%%%%%%%%%%%%

%%% Fig. 1 %%%
\begin{figure*}
  \begin{minipage}{0.49\textwidth}
    \FigureFile(80mm,50mm){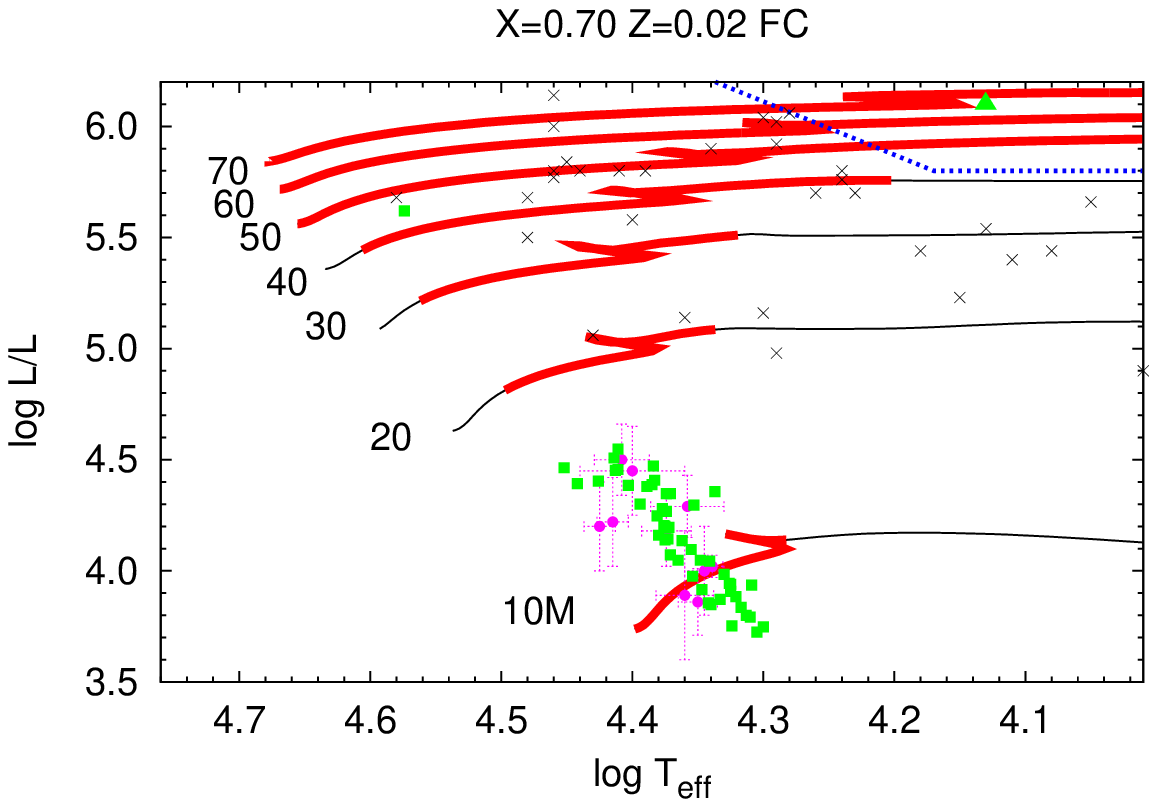}
  \end{minipage}
  \begin{minipage}{0.49\textwidth}
    \FigureFile(80mm,50mm){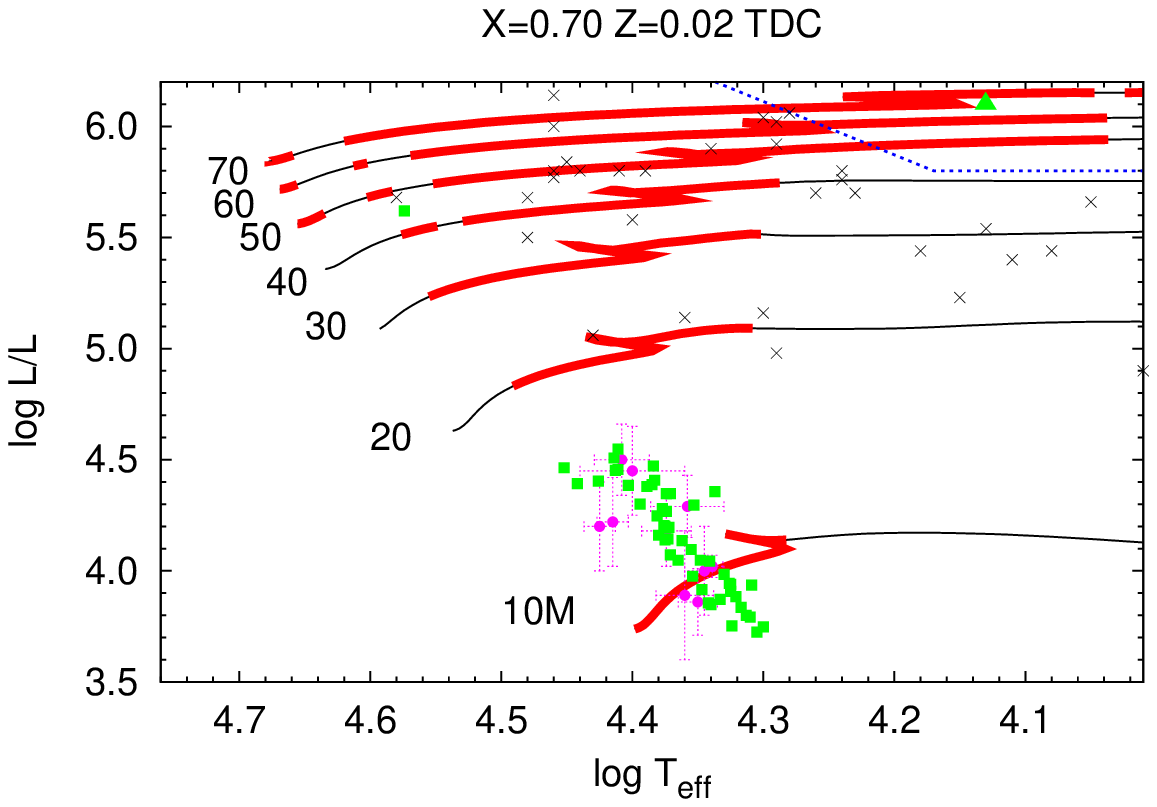}
  \end{minipage}
  \caption{HR diagram showing pulsational instability range for $X=0.70$, $Z=0.02$. The thin solid lines are evolutionary tracks, and the thick parts of the tracks indicate evolutionary stages with pulsational instability. The left and the right panels are results by FC and TDC, respectively. The dashed line is the Humphreys-Davidson (HD) limit. The plots are observed objects. The filled circles and squares are $\beta$ Cep stars listed in \citet{Sterken1993} and Saio, Georgy \& Meynet (\yearcite{Saio2013}), respectively. The filled triangle is HD 50064, which is a candidate for a strange-mode pulsator observed by \citet{Aerts2010}. The crosses are luminous blue variables (LBVs) listed in \citet{vanGenderen2001}.}
  \label{fig:HR_pop1}
\end{figure*}

%%% Fig. 2 %%%
\begin{figure*}
  \begin{minipage}{0.49\textwidth}
    \centering
    \FigureFile(80mm,50mm){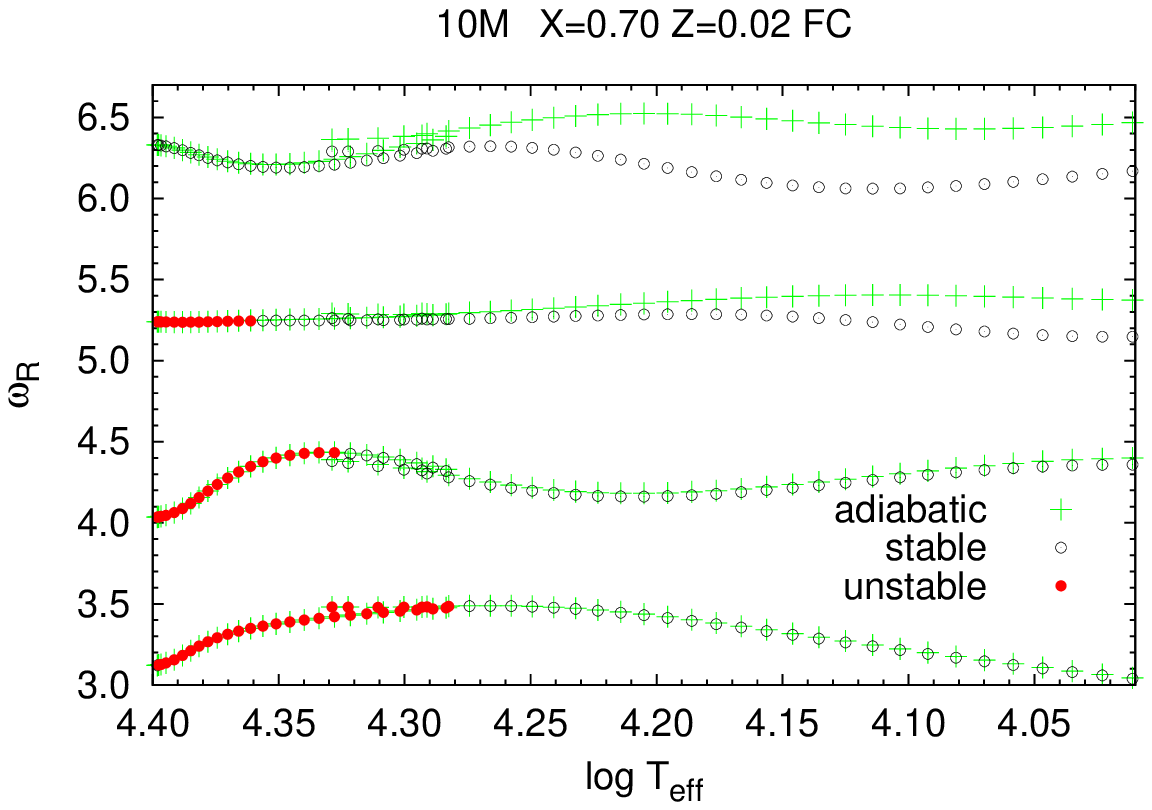}
  \end{minipage}
  \begin{minipage}{0.49\textwidth}
    \centering
    \FigureFile(80mm,50mm){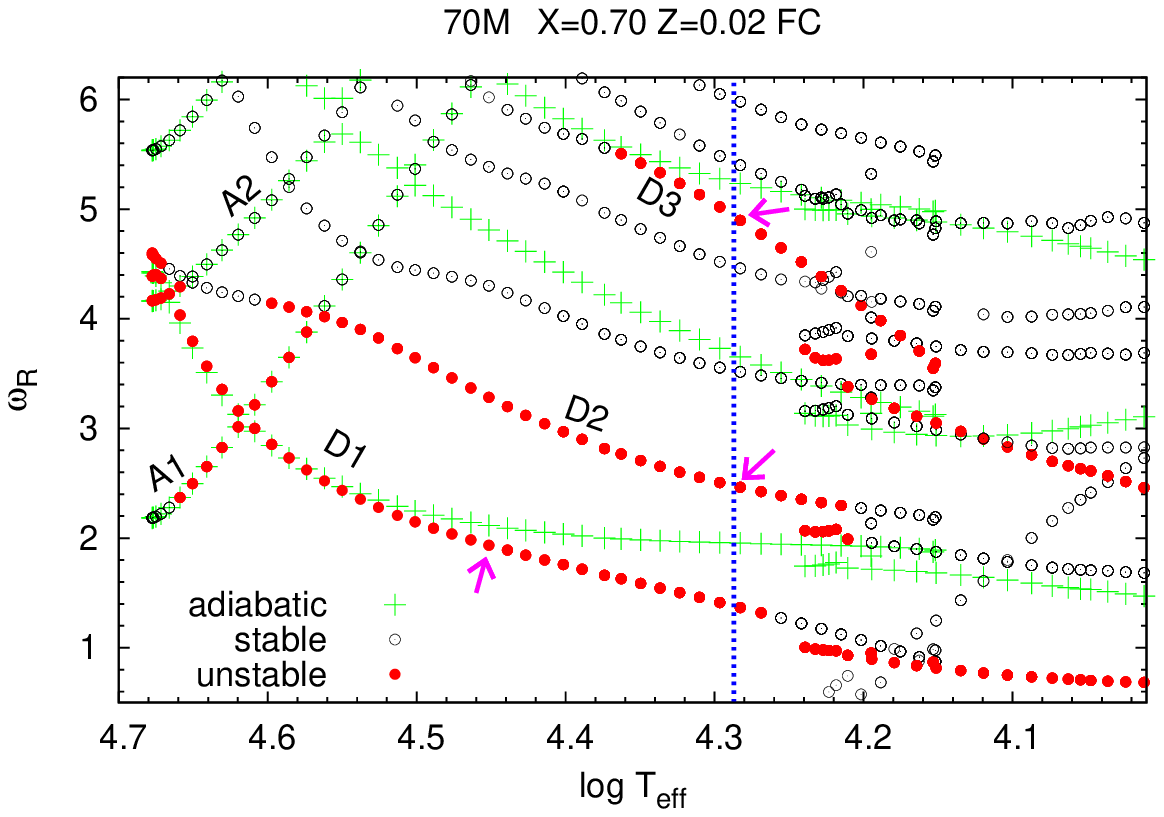}
  \end{minipage}
  \\
  \begin{minipage}{0.49\textwidth}
    \centering
    \FigureFile(80mm,50mm){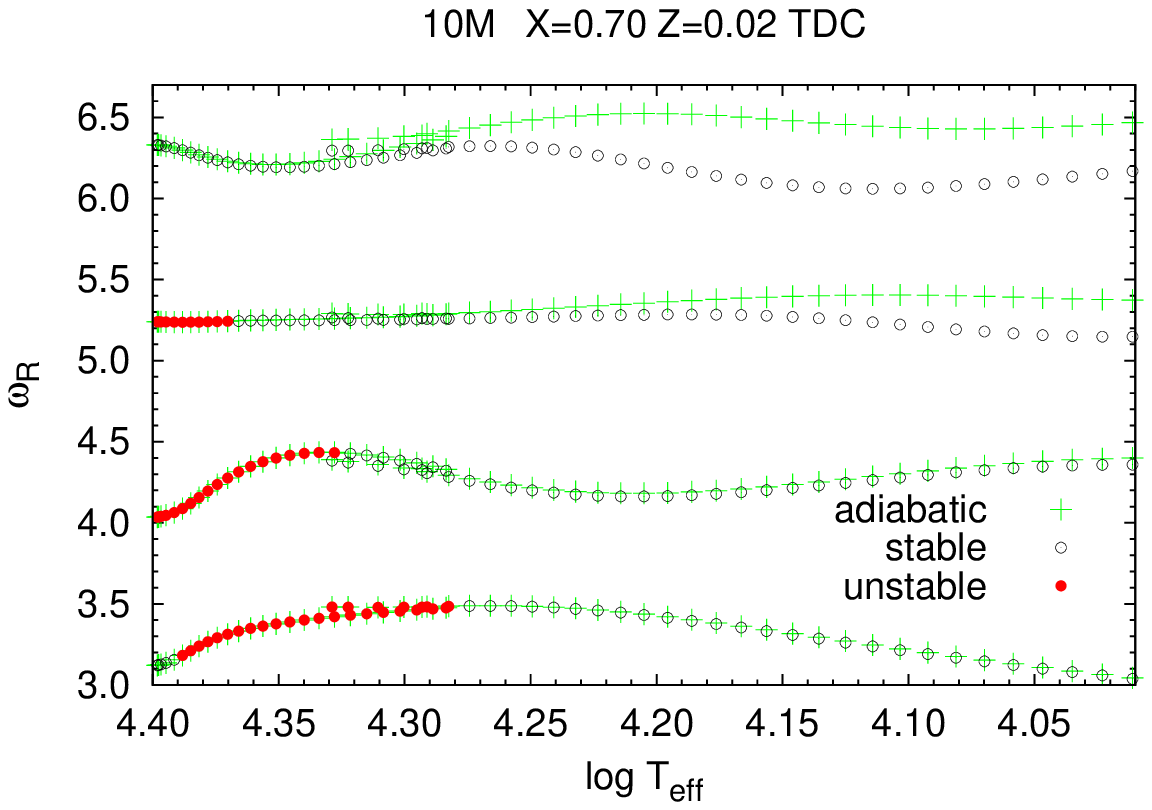}
  \end{minipage}
  \begin{minipage}{0.49\textwidth}
    \centering
    \FigureFile(80mm,50mm){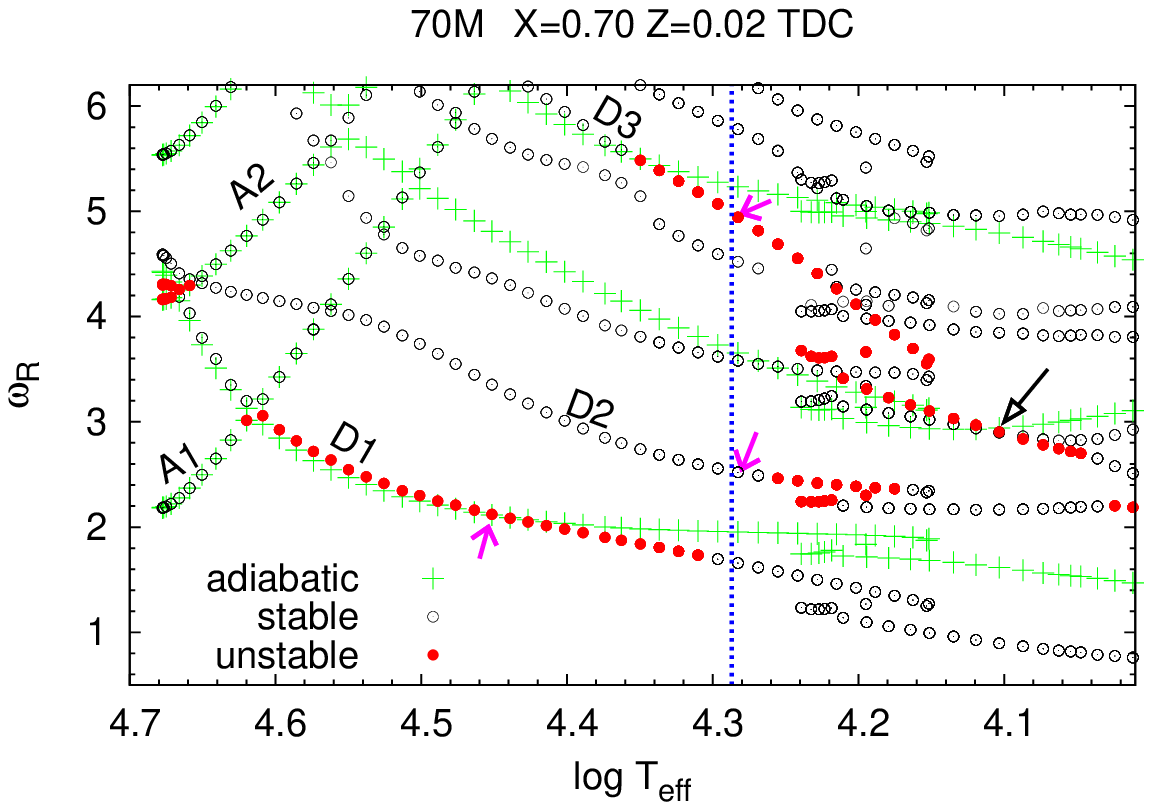}
  \end{minipage}
  \caption{Modal diagrams for $10M_{\odot}$ (left panels) and $70M_{\odot}$ stars (right panels). The top and the bottom panels show results by FC and TDC, respectively. The circles are solutions obtained by the nonadiabatic analysis. The open and the filled circles are stable and unstable modes, respectively. The crosses are solutions of the adiabatic approximation. The vertical dashed lines in the right panels indicate the evolutionary stage crossing the HD limit in the HR diagram. The abscissa is the effective temperature, and the vertical axis is the frequency normalized by multiplying the dynamical time-scale $\tau_{\rm dyn}(\equiv\sqrt{R^3/(GM)})$. Profiles of the 
modes pointed by
%%% ADDED BY FOLLOWING THE REFEREE %%% 
the magenta and the black
%%%%%%%%%%%%%%%%%%%%%%%%%%%%%%%%%%%%%%
arrows are shown in figures
%%% ADDED BY FOLLOWING THE REFEREE %%% 
\ref{fig:work_D1to3}
%%%%%%%%%%%%%%%%%%%%%%%%%%%%%%%%%%%%%% 
and \ref{fig:pop1_vs_3}, respectively.}
  \label{fig:modal_pop1}
\end{figure*}

\section{Comparison between results by FC and TDC}
We first analyze massive stars with $X=0.70$, $Z=0.02$. Here we compare results by FC with zero Lagrangian convective luminosity perturbation $(\delta L_C=0)$ and by TDC. Figure \ref{fig:HR_pop1} shows results of the stability analyses with FC (left) and TDC (right) in the HR diagram. The thick parts on the evolutionary tracks indicate evolutionary stages with pulsational instability. For $M\lesssim 40M_{\odot}$, the instability domain is limited to the high temperature side. For $M\simeq 10-20M_{\odot}$, ordinary modes are destabilized by the $\kappa$-mechanism at the Fe opacity bump, and correspond to $\beta$ Cephei. But the domain extends toward low temperature side for the more massive stars. This tendency is shown both in the results by FC and TDC. But the instability is weaker in the TDC case compared with the FC case. 

Figure \ref{fig:modal_pop1} shows results of the analysis as modal diagrams, of which the abscissa is the effective temperature and the vertical axis is the frequency multiplied by the dynamical time-scale $\tau_{\rm dyn}(\equiv\sqrt{R^3/(GM)})$. While the crosses are the eigenfrequencies obtained by the adiabatic analysis, the circles are the ones by the nonadiabatic analysis. The open and the filled circles correspond to stable and unstable modes, respectively. 

The left panels of figure \ref{fig:modal_pop1} are modal diagrams for the $10M_{\odot}$ star. The top and the bottom panels are results by FC and TDC, respectively. The frequency of each eigenmode does not vary significantly with evolution. We can see that there is a corresponding sequence of adiabatic solutions to each sequence of the nonadiabatic solutions. 

\begin{figure}
  \begin{center}
    \FigureFile(80mm,50mm){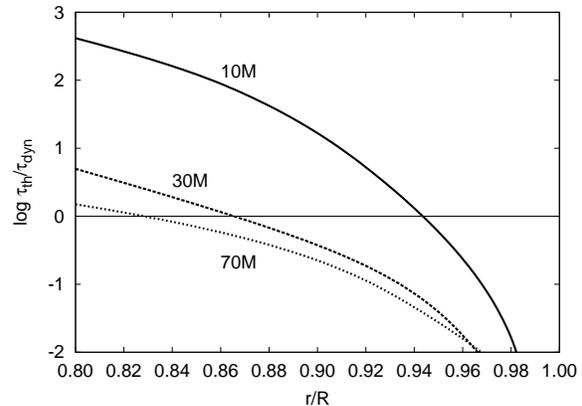}
  \end{center}
  \caption{The ratio of thermal time-scale $\tau_{\rm th}(\equiv\int^R_r4\pi\rho c_pTr^2dr/L)$ to dynamical time-scale $\tau_{\rm dyn}$ for 10, 30 and 70$M_{\odot}$ stars with $\log T_{\rm eff}\simeq 4.2$.}
  \label{fig:tth_over_tdyn}
\end{figure}

As the stellar mass increases, however, the sequences become waving, and come to cross each other complicatedly. For massive stars, there are ascending sequences like ones labeled as A1 and A2, and descending sequences like ones as D1, D2 and D3 in modal diagrams as shown in the right panels of figure \ref{fig:modal_pop1}. The ascending and the descending ones correspond to ordinary modes and strange-modes, respectively. In the right panels, we can see that some of the sequences do not have corresponding adiabatic sequences. This implies that such modes are in quite a different situation from adiabatic pulsations. Figure \ref{fig:tth_over_tdyn} shows profiles of the ratio of thermal to dynamical time-scale for 10, 30 and 70$M_{\odot}$ stars with $\log T_{\rm eff}\simeq 4.2$. With increase in stellar mass, this ratio decreases in a whole star. At the same time, a region with $\tau_{\rm th}/\tau_{\rm dyn}<1$ becomes wider. In such a region, thermal time-scale is shorter than dynamical time-scale, which is comparable to pulsational periods, and then heat transfer takes place much more rapidly unlike in the deep interior, where the thermal time-scale is much longer than the dynamical time-scale. This causes appearance of eigenmodes which cannot be explained by the adiabatic approximation. 

%%% Fig. 3 %%%

\begin{figure}
  \centering
  \FigureFile(80mm,50mm){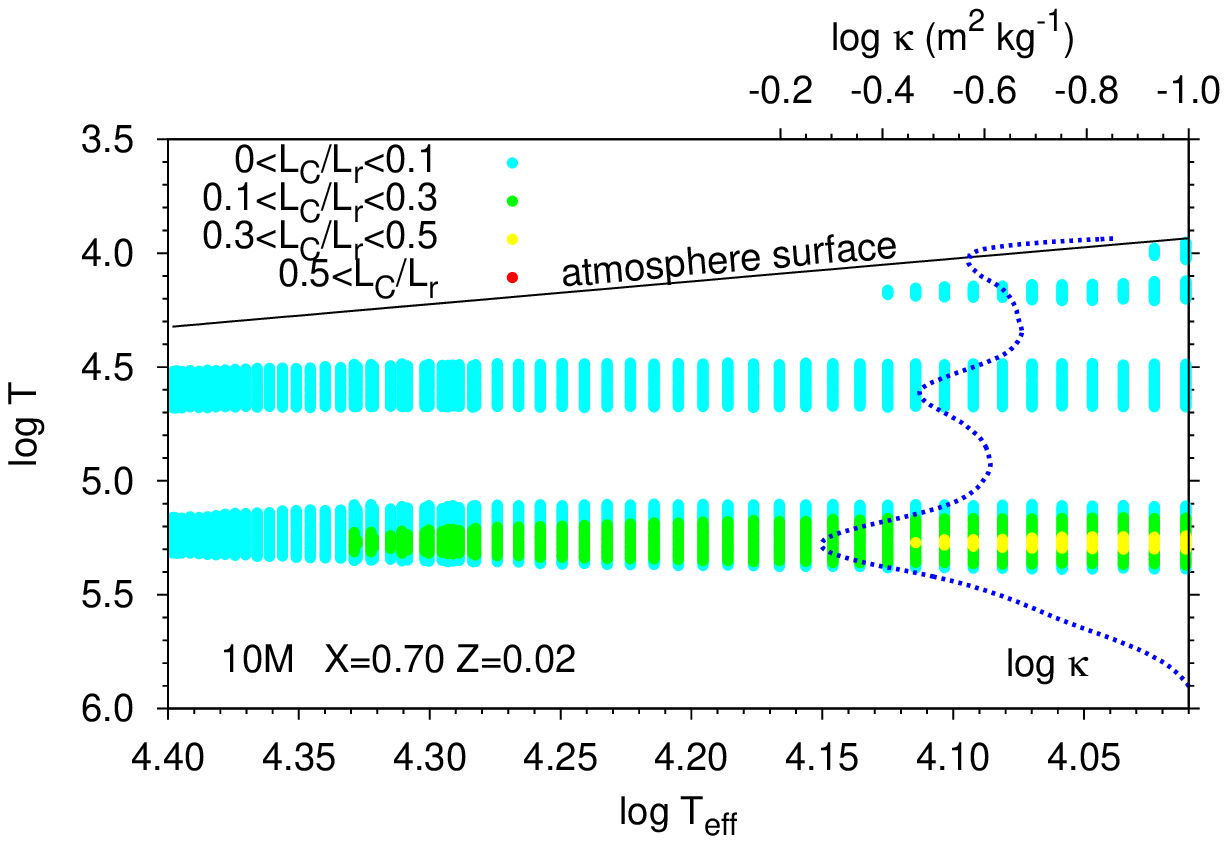}
  \FigureFile(80mm,50mm){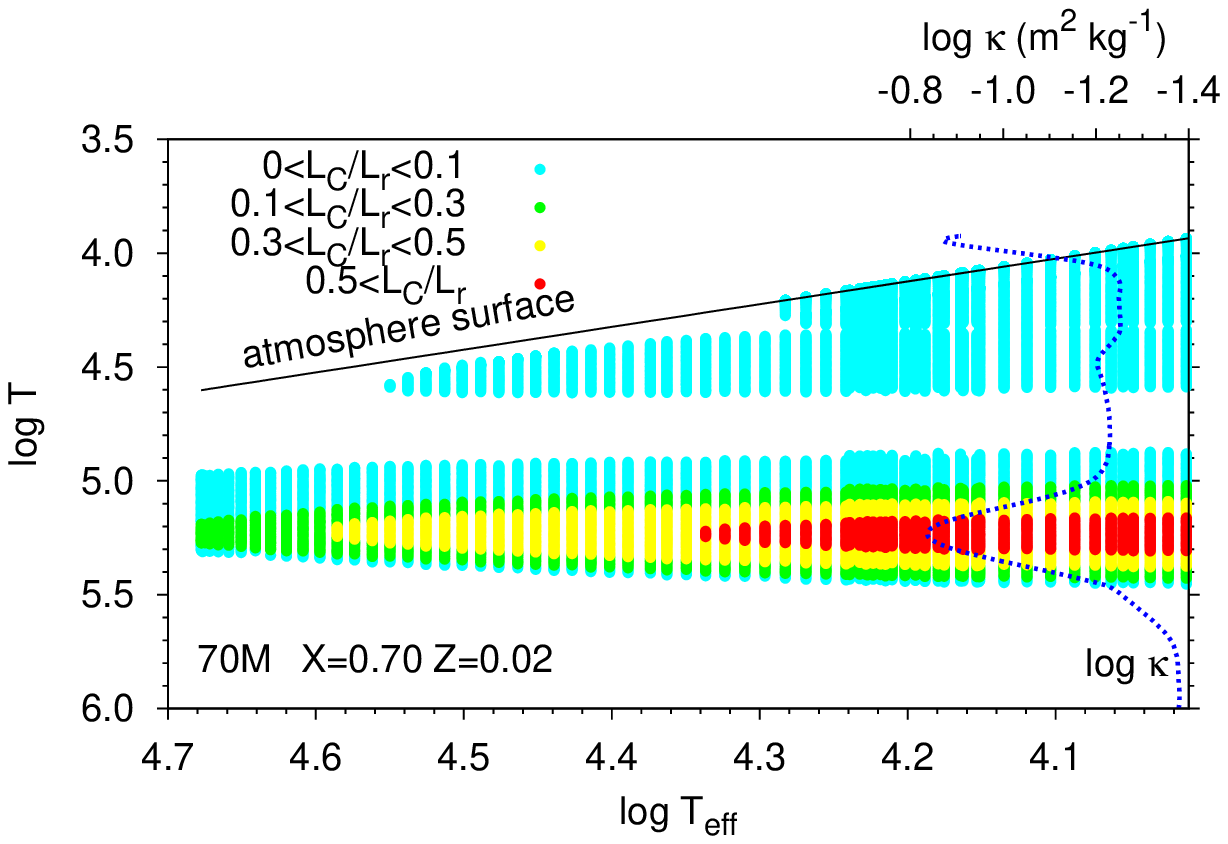}
  \caption{Locations of convection zones for $10M_{\odot}$ (top) and $70M_{\odot}$ stars (bottom). The colors indicate fraction of convective luminosity. The abscissa is the effective temperature, and the vertical axis is the temperature coordinate, i.e., the direction to higher temperature corresponds to that to the stellar center. The dashed line is the opacity profile of the model at $\log T_{\rm eff}\simeq 4$.}
  \label{fig:conv_loc}
\end{figure}

As we can see in figure \ref{fig:modal_pop1}, the degree of the instability hardly varies between results by FC and TDC for the $10M_{\odot}$ stars. For the $70M_{\odot}$ stars, on the other hand, the instability is to some extent weakened with TDC compared with FC. Figure \ref{fig:conv_loc} shows locations of convection zones and contribution of convective luminosity for the $10M_{\odot}$ (top) and the $70M_{\odot}$ sequences (bottom). The convection zones appear due to H, He and Fe opacity bumps. In particular, a convection zone appearing at $\log T=5-5.5$ is generated by the Fe bump, and has a certain contribution of convective luminosity, while convective luminosity is negligible in the other convection zones. For the lower mass stars, the contribution of convective luminosity even in the Fe bump convection zone is so small that convective effects on the pulsational stability is negligible. But as the stellar mass increases, convection comes to considerably contribute to energy transfer, and certainly affects the pulsational stability. 

%%% Fig. 4 %%%

%%% Fig. 5 %%%
\begin{figure}
  \centering
  \FigureFile(80mm,50mm){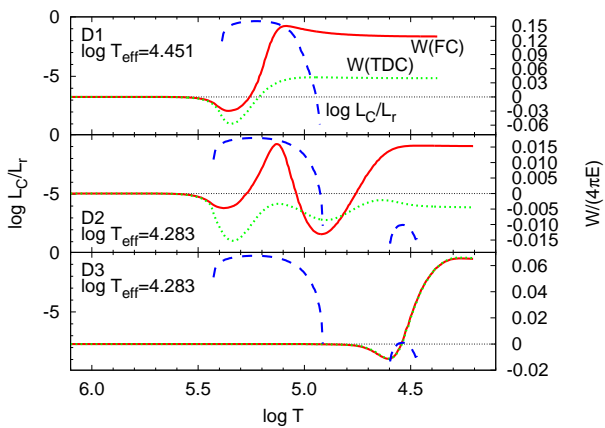}
  \caption{Work integrals of the strange-modes on D1, D2 and D3 sequences for the $70M_{\odot}$ star by FC (solid line) and TDC (short dashed line).
    %%% ADDED BY FOLLOWING THE REFEREE %%%
    These modes are pointed by the magenta arrows in figure \ref{fig:modal_pop1}.
    %%%%%%%%%%%%%%%%%%%%%%%%%%%%%%%%%%%%%%
    Their values are normalized with the total pulsational energy, $E\equiv(\sigma^2_R/2)\int^M_0|\delta{\bf r}|^2dM_r$. The ratio of convective to total luminosity is also shown as the long dashed line. The corresponding evolutionary stages are indicated with the effective temperature labeled in each panel.}
  \label{fig:work_D1to3}
\end{figure}

Figure \ref{fig:work_D1to3} shows work integrals of the strange-modes on D1, D2 and D3 sequences for the $70M_{\odot}$ star. With FC, the strange-modes on D1 and D2 sequences are excited around the Fe bump convection zone. On the other hand, the one on D3 is not excited there, but around the He bump convection zone. As for D1 and D2, damping works more strongly with TDC than with FC, since convective luminosity certainly contributes in the Fe bump convection zone. That is why the instability is weaker with TDC than with FC as shown in figure \ref{fig:modal_pop1}. The ordinary modes on A1 and A2 sequences are also excited at the Fe bump convection zone with FC. The instability of A1 is weakened with TDC like D1 and D2, while that of A2 does not vary between FC and TDC since it occurs at the beginning of the main-sequence stage, at which contribution of convective luminosity is relatively small compared with evolutionary stages with lower effective temperature. The results for the D3 mode are also almost identical with FC and TDC, since the contribution of convective luminosity is negligible in the He bump convection zone. 

\citet{Glatzel1996} analyzed pulsational stability of massive stars through their evolutions like our study. They performed a nonadiabatic analysis by adopting two types of FC with zero Lagrangian and Eulerian perturbations of convective luminosity ($\delta L_C=0$ and $L'_C=0$). In their results, the instability of D1 and D2 is suppressed with $L'_C=0$, while it appears with $\delta L_C=0$ as much as in our study. But the instability of D3 invariably exists both with the two types of FC, and is independent of the treatment of convection similarly to our study.

\section{Excitation mechanism}
The ordinary modes (A1 and A2 sequence modes) are excited by the $\kappa$-mechanism at the Fe bump. Unstable strange-modes with adiabatic counterparts (D1 sequence modes in the main-sequence stage) are also excited by the $\kappa$-mechanism (Saio et al. \yearcite{Saio1998}). On the other hand, unstable strange-modes without adiabatic counterparts (D2 and D3 sequence modes) are excited by another mechanism. The instability of the latter type of strange-mode has been called ``strange-mode instability'' in previous studies. \citet{Glatzel1994} and \citet{Saio1998} suggested that dominance of radiation pressure and a large phase lag between density and pressure perturbations are important for the strange-mode instability. Here we discuss the phase lag in the excitation of the latter type of strange-mode.

%%% Fig. 6 %%%
\begin{figure}
  \centering
  \FigureFile(80mm,50mm){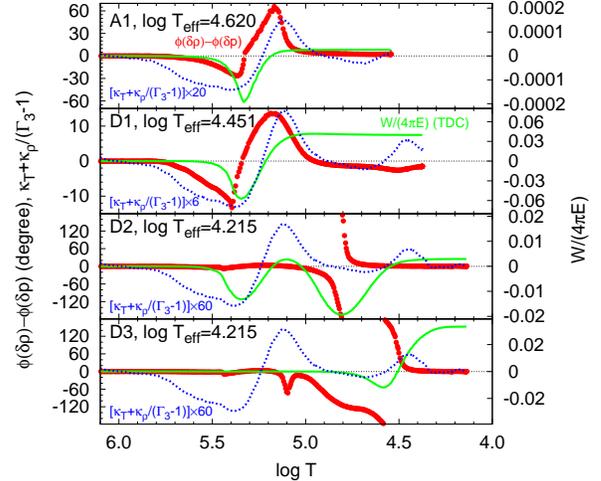}
  \caption{Work integral (solid line) and phase lag between density and pressure perturbations (dot) for the unstable ordinary mode on the sequence A1, and for the unstable strange-modes on the sequences D1 to D3. The values of the work integrals are normalized in the same way as in figure \ref{fig:work_D1to3}. These modes are computed with TDC. They are for the $70M_{\odot}$ models, and the corresponding evolutionary stages are indicated with the effective temperature labeled in each panel. This figure also shows profiles of $\kappa_T+\kappa_\rho/(\Gamma_3-1)$ (dashed line). Roughly speaking, the $\kappa$-mechanism should take place in a region with $d[\kappa_T+\kappa_\rho/(\Gamma_3-1)]/d\log T<0$.}
  \label{fig:phase_D1toD3}
\end{figure}

Figure \ref{fig:phase_D1toD3} shows work integrals and profiles of phase lags for an unstable ordinary mode on A1 sequence, and for unstable strange-modes on D1 to D3 sequences. The top panel shows the unstable ordinary mode appearing on A1 sequence. This mode is excited by the $\kappa$-mechanism at the Fe bump. As shown in the figure, the excitation zone satisfies the analytically derived condition for the occurrence of the $\kappa$-mechanism \citep{Unno1989}: 
\begin{eqnarray}
  \frac{d}{d\log T}\left[\kappa_T+\frac{\kappa_\rho}{\Gamma_3-1}\right]<0,
  \label{eq:k-mech}
\end{eqnarray}
where $\kappa_T\equiv (\partial\ln\kappa/\partial\ln T)_\rho$, $\kappa_\rho\equiv (\partial\ln\kappa/\partial\ln\rho)_T$ and $\Gamma_3-1\equiv (\partial\ln T/\partial\ln\rho)_S$. That is true also for the D1 mode. The phase lag between the density and the pressure Lagrangian perturbations, $\phi(\delta\rho)-\phi(\delta p)$, becomes positive in an excitation zone, while negative in a damping zone. In both cases of A1 and D1, damping takes place just inside the excitation zone. At the transition point from the damping to the excitation, the phase lag is zero. The phase lag has its maximum in the middle of the excitation zone.

%%% Fig. 7 %%%
\begin{figure}
  \begin{center}
    \FigureFile(80mm,50mm){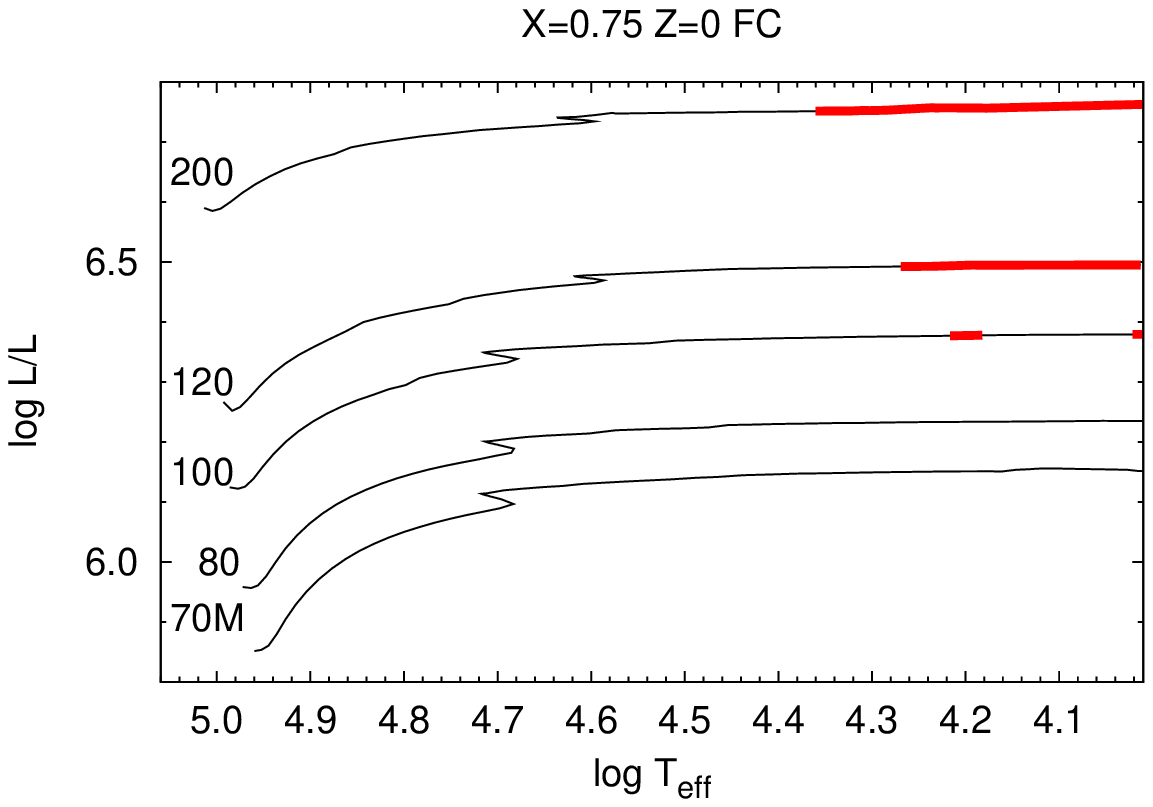}
    \FigureFile(80mm,50mm){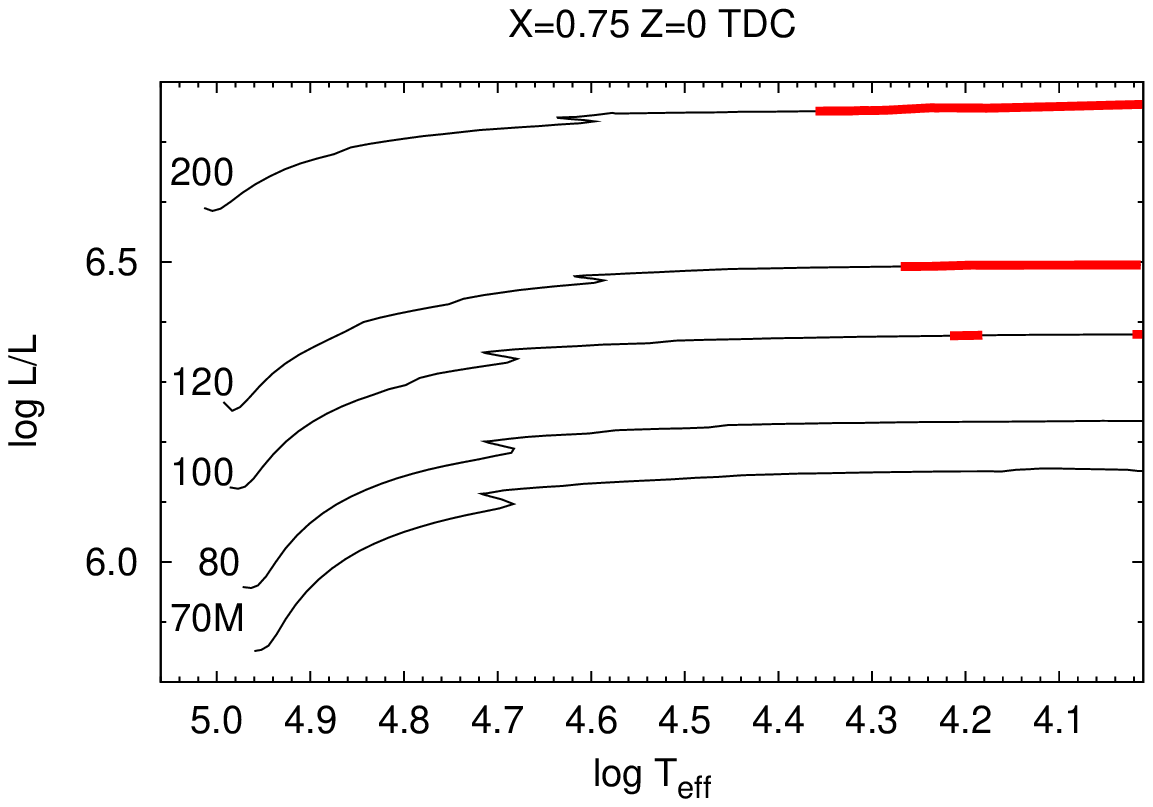}
  \end{center}
  \caption{HR diagrams showing instability regions for $X=0.75$, $Z=0$. The top and the bottom panels are results by FC and TDC, respectively. The definitions of lines are the same as figure \ref{fig:HR_pop1}.}
  \label{fig:HR_pop3}
\end{figure}

On the other hand, the excitation of the D2 and the D3 modes, which are strange-modes without adiabatic counterparts, is different from the above. The D3 mode is excited around the He bump. But the excitation zone extends outside of the zone satisfying equation (\ref{eq:k-mech}) ($\log T\simeq 4.2-4.4$). This implies that the excitation should not be the $\kappa$-mechanism. Similarly to the cases of A1 and D1, damping takes place just inside of the excitation zone. But the phase lag becomes $180^{\circ}$ at the transition point unlike the above cases. In the excitation zone, the phase lag varies from $180^{\circ}$ to 0. That is, the phase lag can become much larger in the strange-mode instability than in the case of the $\kappa$-mechanism. As for the D2 mode, the excitation takes place at the Fe bump ($\log T\simeq 5.1-5.3$) and between the Fe and the He bumps ($\log T\simeq 4.6-4.8$). The excitation at the Fe bump occurs in the zone satisfying equation (\ref{eq:k-mech}), and can be regarded as the $\kappa$-mechanism. On the other hand, the one between the Fe and the He bumps happens in a zone not related to any opacity bumps, and the profile of the phase lag is the same type as that of the D3 mode. That is, the $\kappa$-mechanism excitation and the strange-mode instability act together on the D2 mode. 

%%% Fig. 8 %%%
\begin{figure}
  \begin{center}
    \FigureFile(80mm,50mm){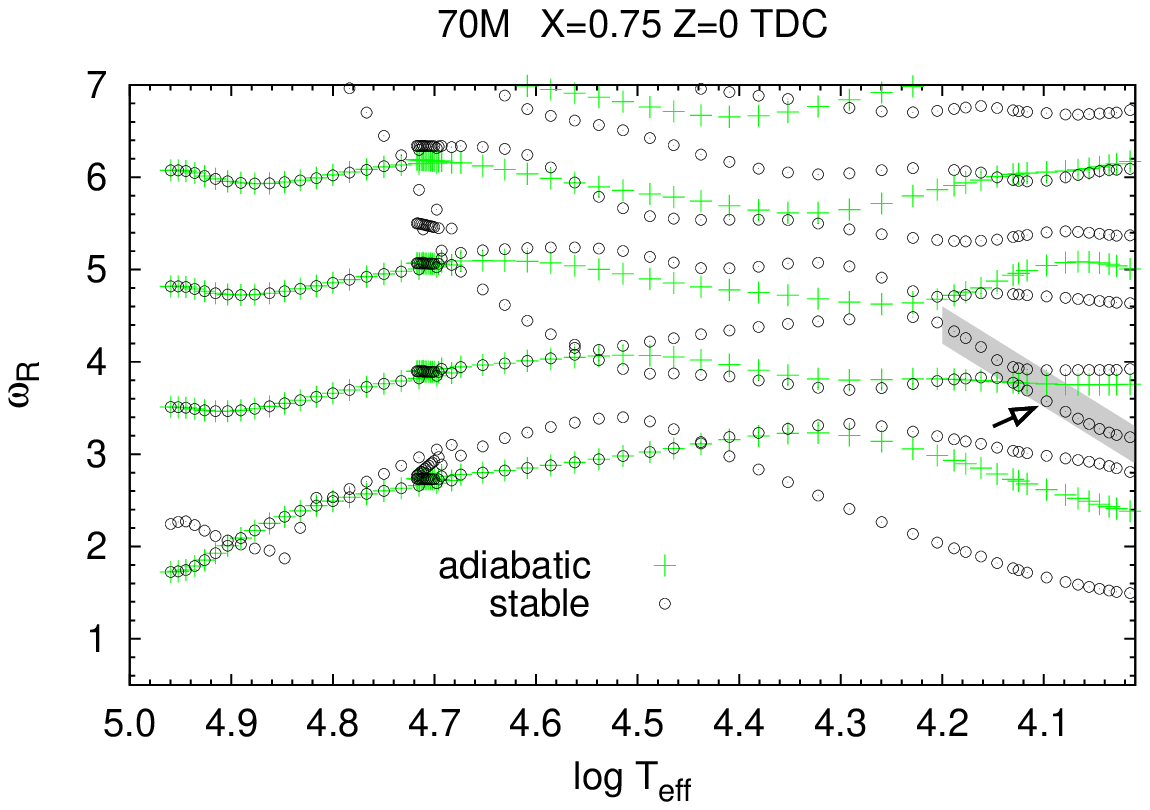}
    \FigureFile(80mm,50mm){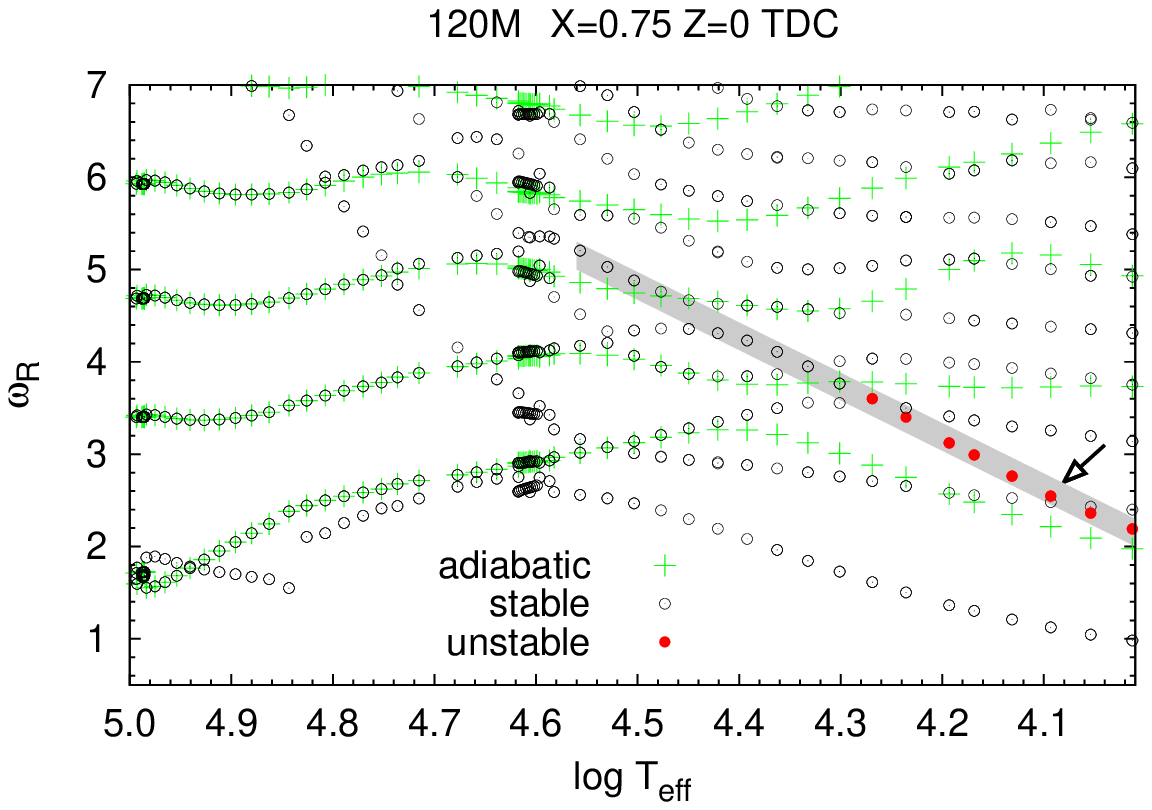}
  \end{center}
  \caption{Modal diagrams by TDC for $70M_{\odot}$ (top) and $120M_{\odot}$ stars (bottom) with $X=0.75$, $Z=0$. The definition of the symbols are the same as in figure \ref{fig:modal_pop1}. The steeply descending sequence corresponding to D3 sequence in the $Z=0.02$ case is highlighted. Profiles of the modes pointed by the arrows on the highlighted sequences are shown in figure \ref{fig:pop1_vs_3}.}
  \label{fig:modal_pop3}
\end{figure}

\section{Comparison with the zero-metallicity case}
We also carry out a stability analysis for zero-metallicity models with Population III composition, $X=0.75$, $Z=0$. Figure \ref{fig:HR_pop3} shows the instability domains by FC and TDC for the zero-metallicity case. Unlike the metal-rich case, the pulsational instability is limited to the post-main sequence stage of very massive stars with $M\gtrsim 100M_{\odot}$. The results by FC and TDC are almost identical, since the unstable modes are excited at the He bump convection zone, where convective luminosity is negligible. Figure \ref{fig:modal_pop3} shows modal diagrams obtained by TDC for the $70M_{\odot}$ (top) and the $120M_{\odot}$ sequences (bottom). Also in the $Z=0$ case, a steeply descending sequence like D3 of the $Z=0.02$ case, highlighted in figure \ref{fig:modal_pop3}, appears in the low temperature side. For $M\lesssim 100M_{\odot}$, no unstable mode is found. But unstable modes appear on this type of sequence for the more massive stars. The mode is destabilized through the strange-mode instability around the He bump similarly to the D3 mode. On the other hand, the ordinary modes and the strange-modes corresponding to D1 and D2 are never destabilized since the Fe bump does not exist.

\begin{figure}
  \begin{center}
    \FigureFile(80mm,50mm){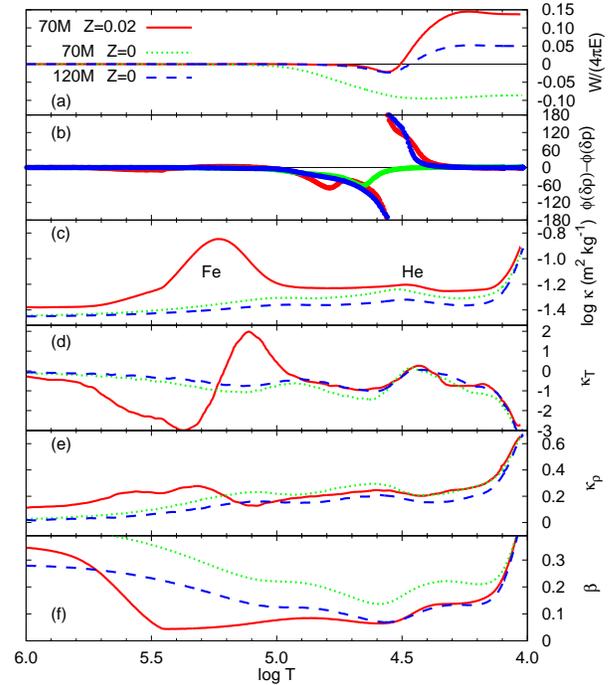}
  \end{center}
  \caption{Profiles of the strange-modes on the steeply descending sequences, pointed by the arrows in the modal diagrams in Figs. \ref{fig:modal_pop1} and \ref{fig:modal_pop3} (a,b), and those of the corresponding equilibrium models for $70M_{\odot}$ with $X=0.70$, $Z=0.02$ and for $70M_{\odot}$ and $120M_{\odot}$ with $X=0.75$, $Z=0$ at $\log T_{\rm eff}\simeq 4.1$ (c--f). From top to bottom, work integral normalized with total pulsation energy $W/(4\pi E)$ (a), phase lag between density and pressure perturbations $\phi(\delta\rho)-\phi(\delta p)$ in degree (b), opacity $\kappa$ (c), first derivative of opacity with respect to temperature $\kappa_T$ (d), that with respect to density $\kappa_\rho$ (e), and ratio of gas pressure to total pressure $\beta$ (f). Table \ref{tab:pop1_vs_3} shows the characteristics of the modes and the equilibrium models shown here.}
  \label{fig:pop1_vs_3}
\end{figure}

Although the excitation of the unstable modes takes place around the He bump, stability of these modes actually depends on metallicity. Figure \ref{fig:pop1_vs_3} compares $70M_{\odot}$ models with $Z=0$ and $Z=0.02$, which have almost the same luminosity and effective temperature. Since the amplitude of strange-modes is confined to the outer region, the properties of the model envelope are important. These are determined by the stellar mass, the chemical composition, the effective temperature and the luminosity. As shown in Table \ref{tab:pop1_vs_3}, all these values are almost the same between the $70M_{\odot}$ models with $Z=0$ and $Z=0.02$. However, the stability is different between them. In figure \ref{fig:pop1_vs_3}, the opacity derivatives with respect to temperature and density does not seem so different around the He bump between the $Z=0$ and the $Z=0.02$ cases. On the other hand, the Fe bump exists in the $Z=0.02$ case, while it does not in the $Z=0$ case. Radiation pressure becomes strong and dominant around an opacity bump in order to transfer energy radiatively against the large opacity. As shown in figure \ref{fig:pop1_vs_3}, the ratio of gas to total pressure $\beta$ is much lower at the Fe bump with $Z=0.02$ than with $Z=0$. Then, the existence of the Fe bump leads to dominance of radiation pressure in the whole envelope, and to the strange-mode instability. The second panel of figure \ref{fig:pop1_vs_3} shows that the phase lag between density and pressure perturbations varies from 0 to $180^{\circ}$ in the excitation zone for the unstable modes similarly to the D3 mode. 

%%% Table 1 %%%
\begin{table*}
  \caption{Characteristics of modes and equilibrium models shown in figure \ref{fig:pop1_vs_3}.}
  \begin{center}
  \label{tab:pop1_vs_3}
  \begin{tabular}{cccccc}\hline
    $M$ & $Z$ & $\log T_{\rm eff}$ & $\log L$ & Period & $e$-folding time\footnotemark[$*$] \\
    ($M_{\odot}$) & & (K)           & ($L_{\odot}$) & (d) & (yr) \\ \hline
    70 & 0.02 & 4.103             & 6.149    & 18.5   & 0.550 \\
    70 & 0    & 4.097             & 6.156    & 18.9   & $-$0.754 \\
    120 & 0   & 4.093             & 6.495    & 31.3   & 2.586 \\ \hline
    \multicolumn{6}{@{}l@{}}{\hbox to 0pt{\parbox{85mm}{\footnotesize
          \par\noindent
          \footnotemark[$*$] The positive and the negative values correspond to growth and damping of amplitude, respectively.
        }\hss}}
  \end{tabular}
  \end{center}
\end{table*}

Even in the zero-metallicity case, however, the strange-mode instability takes place for $M\gtrsim 100M_{\odot}$. As the stellar mass increases, radiation pressure becomes dominant in the envelope due to increase in luminosity. The $120M_{\odot}$ case in figure \ref{fig:pop1_vs_3} corresponds to this. The phase lag varies from 0 to $180^{\circ}$ in the excitation zone similarly to the case of $70M_{\odot}$, $Z=0.02$.

%%% Fig. 9 %%%

\section{Discussion}
The previous studies have found unstable strange-modes in hot massive stars with FC. Especially, \citet{Kiriakidis1993} suggests the association with the HD limit phenomenon. As shown in figure \ref{fig:HR_pop1}, LBVs appear in the higher temperature side of the HD limit. They experience sporadic mass eruptions, and the instability of the strange-modes has been thought to be related to them. Although the previous studies have adopted the FC approximation, the present study revealed that the instability certainly appears even with taking into account effects of convection. Although we carry out a linear analysis in this study, a nonlinear analysis is required to know if the instability can indeed induce the mass eruptions.

Some of LBVs are distributed in the lower effective temperature side region below the horizontal part of the HD limit line, which corresponds to the post-main sequence stage of $20-40M_{\odot}$ stars. In this region, no pulsational instability is found. In this study, calculation of stellar evolution is carried out without taking into account mass loss, and stopped before reaching the red supergiant stage. According to calculations with mass loss (e.g. \cite{Ekstrom2012, Georgy2012}), on the other hand, evolutionary models in such a mass range first evolve toward red supergiants, and lose substantial mass. After that, they evolve back toward blue supergiants (blue loop stage). \citet{Saio2013} showed that models in the blue loop stage have high $L/M$ ratio and present unstable strange-modes even if no unstable strange-modes appear while evolving toward red supergiants. 
    
When we consider stars experiencing mass loss, we should be aware of effects of atmosphere states. \citet{Godart2011} found a difference in stability of strange-modes between with adoptions of an Eddington-grey atmosphere model and of a dynamic atmosphere model, FASTWIND \citep{Puls2005}. Further investigations are worth doing in combination with TDC.

We also carry out a stability analysis for Population III models. The Population III stars might play an important role in the chemical evolution of the early Universe. Since they have no or few heavy elements, very massive stars could be formed due to lack of cooling by heavy element emission lines in the star formation stage (e.g. Bromm et al. \yearcite{Bromm1999}; Abel et al. \yearcite{Abel2002}; \authorcite{Omukai2003} \yearcite{Omukai2003}). In the mass range of $130-300M_{\odot}$, a star is thought to evolve toward a pair-instability supernova (PISN), which provides significantly different chemical composition from that by a core-collapse supernova. But the existence of PISN is controversial, and recent detailed comparisons between the observations of extremely metal poor (EMP) stars \citep{Cayrel2004} and the nucleosynthesis yields of PISN models \citep{Umeda2002, Heger2002} have shown that the PISN yields are not suitable to the abundance patterns of EMP stars. The instability of the strange-mode might be inhibitory to the evolution toward PISN. This is in agreement with the above results. 

\section{Conclusion}
We carry out a nonadiabatic analysis of strange-modes with TDC for hot massive stars. Compared with results by FC, the instability is weaker for modes excited at the Fe bump. Convection certainly contributes to energy transfer around the Fe bump, and gives damping effects on pulsations. In spite of this, we confirm that instability of strange-modes certainly remains in hot massive stars even with taking into account TDC. 

We also carefully examine properties of the strange-mode instability, which acts on strange-modes without adiabatic counterparts. Unlike the case of the $\kappa$-mechanism, the phase lag between density and pressure perturbations varies from 0 to $180^{\circ}$ in an excitation zone. Besides, we confirm by comparing the models with $Z=0$ and $Z=0.02$ that dominance of radiation pressure is important for the strange-mode instability. These results are in agreement with the previous analytic works of \citet{Glatzel1994} and \citet{Saio1998}.

Since the growth time-scale of strange-modes is extremely short, nonlinear phenomena are expected to take place. Then, nonlinear analyses are worth doing especially to understand associations with LBV phenomena and effects on evolution of Population III very massive stars toward PISN.

\bigskip

The authors are grateful to the anonymous referee for useful comments to improve our manuscript. They would like to thank M\'elanie Godart for her helpful advices and comments that led to the improvement of the original manuscript. They also thank Hideyuki Saio for his useful comments about strange-modes, and Takashi Sekii, Masao Takata, Othman Benomar and Kazuhiro Maeda for fruitful discussions. This study is financially supported by Japan Society for the Promotion of Science Grant-in-Aid for Research Fellows.

\end{document}